\documentclass{article}
\usepackage{emulateapj,onecolfloat,graphics}

\begin{document}
\twocolumn[
\title{Constraining evolution in the Halo Model using galaxy
redshift surveys}
\author{Renbin Yan$^{1}$, Darren S. Madgwick$^{1,2}$
 \& Martin White$^{1,2,3}$}
\affil{$^1$ Department of Astronomy, University of California,
Berkeley, CA 94720}
\affil{$^2$ Lawrence Berkeley National Laboratory, Berkeley, CA 94720}
\affil{$^3$ Department of Physics, University of California,
Berkeley, CA 94720}

\begin{abstract}
We use the latest observations from the 2dF Galaxy Redshift Survey to
fit the conditional luminosity function (CLF) formulation of the
halo-model for galaxies at $z=0$.  This fit is then used to test
the extent of evolution in the halo occupation distribution (HOD)
to $z=0.8$, by
comparing the predicted clustering from this CLF to preliminary 
results from the DEEP2 Redshift Survey.
We show that the current observations from the DEEP2 Redshift Survey
are remarkably consistent with no evolution in the CLF from $z=0$ to
$z=0.8$.  This result is surprising, in that it suggests that there
has been very little change in the way galaxies occupy their host dark
matter halos over half the age of the Universe.  We discuss in detail
the observational constraints we have adopted and also the various
different selection effects in each survey and how these impact on the
galaxy populations encountered in each survey.
\end{abstract}

       ]

\section{Introduction}

Despite impressive advances in many areas over the last decade, galaxy
formation remains one of the central unsolved puzzles in cosmology.
While it is now generally accepted that galaxies form within dark matter
halos (White \& Rees \cite{WhiRee}), whose properties we know about with
increasing reliability, the details of this process are only poorly
understood.
With the advent of large-scale surveys of galaxies and ever more
sophisticated computer simulations of structure formation we need to develop
a more nuanced view of galaxy formation than has been typical until now.
Key to advances in this subject is a framework within which to compare
theory and observations in as transparent a way as possible.

While studies of individual galaxies can shed light on many of the relevant
physical processes which shape them, galaxy clustering stands as the most
successful route to constraints on galaxy formation models to date.  Thus
we desire a framework within which we can interpret the numerous measurements
of galaxy clustering that have accumulated over the years.

The `halo model' provides such a framework.  Building on the insights of
semi-analytic galaxy formation
(e.g.~Kauffmann, White \& Guiderdoni \cite{KauWhiGui};
 Cole et al.~\cite{CAFNZ};
 Somerville \& Primack \cite{SomPri})
and high-resolution hydrodynamic simulations including star-formation
and feedback
(Katz, Hernquist \& Weinberg~\cite{KatHerWei};
 Gardner et al.~\cite{GKHW};
 Pearce et al.~\cite{Pearce};
 White, Hernquist \& Springel~\cite{WhiHerSpr};
 Yoshikawa et al.~\cite{YTJS})
the halo model has been extensively developed in recent years,
e.g.~Jing, Mo \& Borner.~\cite{JinMoBor};
 Benson et al.~\cite{BCFBL};
 Seljak~\cite{Sel00};
 Peacock \& Smith~\cite{PeaSmi};
 Ma \& Fry~\cite{MaFry};
 Scoccimarro et al.~\cite{SSHJ};
 White \cite{Whi01};
 Scoccimarro \& Sheth~\cite{ScoShe};
 Berlind \& Weinberg \cite{BW02};
 Scranton \cite{Scr02};
 Yang, Mo \& van den Bosch \cite{YMvdB};
 (see Cooray \& Sheth \cite{CooShe} for a recent review and references to
 the earlier literature).

The halo model postulates that all galaxies lie in virialized halos and that
the number and type of galaxies in halos is determined primarily (or entirely)
by the halo mass.
Knowledge of the number of galaxies as a function of halo mass, known as the
halo occupation distribution (HOD), their spatial and velocity distribution
within the halos and a model for the spatial clustering of dark matter halos
is sufficient to predict most observables in large-scale structure.
The halo model provides a new way of thinking about galaxy bias which is more
physically informative than the earlier schemes.
It also provides a conceptual division between `galaxy formation' and
`cosmology' in that the latter affects the spatial distribution and number
of dark matter halos, while the former describes the properties and number
of galaxies which form within the halos.
The key advantage of having this conceptual separation between
halo and galaxy properties is that
the evolution in the spatial distribution and number of dark matter
halos can be well understood through the use of numerical simulations,
allowing us to exclusively constrain galaxy formation processes with the
use of recent large galaxy redshift surveys. 

The recent completion of the 2dF Galaxy Redshift Survey (2dFGRS,
Colless et al.~\cite{2dF}), together with
the ongoing progress of the Sloan Digital Sky Survey (SDSS, Strauss et
al.~\cite{Strauss}), 
have heralded a new age of precision quantification in the
galaxy population in the local ($z\sim0$) Universe.  The sheer size of
these data sets enables a particularly accurate characterization
of the properties of the galaxies themselves, together with how these
galaxies are distributed with respect to each other.  These two ingredients
will be key to resolving many outstanding issues in galaxy formation
and evolution.
The DEEP2 Galaxy Redshift Survey (Davis et al.~\cite{DEEP2}) will herald a
similar degree of improvement in our understanding of the galaxies
at $z\sim1$.  The goal of this paper is to
link these two significant advances in our understanding of the galaxy
population, by performing an initial comparison between high
fidelity 2dFGRS results 
and preliminary DEEP2 clustering results (Coil et al.~\cite{coil}) 
within the halo-model framework.

We use the latest observations from the 2dFGRS to constrain the HOD at
$z\sim0$, by adopting the conditional luminosity function (CLF) formalism
developed by Yang et al.~(\cite{YMvdB}).
We make use of a Markov-Chain Monte-Carlo (MCMC; see
Gilks, Richarson \& Spiegelhalter \cite{gilk})
procedure to explore the multi-dimensional parameter space of the CLF.  
At present we consider the cosmology to be fixed.  Specifically we assume
a $\Lambda$CDM model with $\Omega_{\rm m}=0.3$, $\Omega_\Lambda=0.7$,
$H_0=100\,h\,{\rm km}\,{\rm s}^{-1}{\rm Mpc}^{-1}$ with $h=0.7$, 
$\Omega_{\rm B}h^2=0.02$, $n=0.95$ and $\sigma_8=0.9$
(close to the best fit to the CMB data for this cosmology).
We note in passing that the matter power spectrum is insensitive to the
precise value of $\Omega_{\rm B}h^2$ assumed and that the normalization
is well fixed by the recent WMAP data (Bennett et al.~\cite{WMAP}) once
a specific cosmology is chosen.

Once we have constrained the CLF at $z\sim0$ in this way, we proceed
to test the extent of evolution in the galaxy formation processes
which are quantified by this function.  In particular (by assuming a
given cosmological model in our N-body simulations) we use this CLF
to make predictions about observable galaxy properties at $z\sim1$ --
which are tested using the recent clustering observed in the DEEP2 Galaxy
Redshift Survey (Coil et al.~\cite{coil}). 

The outline of this paper is as follows:  In \S\ref{sec:clf} we describe
the conditional luminosity function formulation of the halo model and
summarize the main parameters in this model.  Section \ref{sec:constraints}
gives a detailed description of all the observational constraints we will
use in our analysis, and in particular describes how we deal with the
different selection effects in the two surveys.  
Our fitting procedure is outlined in \S\ref{sec:MCMC}, which also discusses
the best-fitting parameters output by the MCMC.
In \S\ref{sec:results} we present our prediction for the clustering we
would expect to see in the DEEP2 Redshift Survey if the CLF hadn't
undergone any evolution, and then compare this to what is actually
observed and we present our conclusions in \S\ref{sec:conclusions}.

\section{The Conditional Luminosity Function} \label{sec:clf}

Here we briefly review the conditional luminosity function (CLF) formalism
of Yang et al.~(\cite{YMvdB}).
Further details on the halo model and references to the literature can be
found in Appendix \ref{sec:clfappendix}.

Typically when implementing the halo-model, the HOD is assumed to
simply yield the probability of having $N$ galaxies in a halo of mass
$M$, i.e.~$P(N|M)$, and is hence independent of the individual
properties of the galaxies under consideration
(e.g.~Seljak \cite{Sel00}; Peacock \& Smith \cite{PeaSmi}).
A large advance in the development of the HOD function was the recent work of
Yang et al.~(\cite{YMvdB}), who focussed attention on the conditional
luminosity function, $\Phi(L|M)$: the luminosity function
of galaxies in halos of mass $M$.
The CLF extends the HOD by treating galaxies not as indistinguishable objects
but as carrying a luminosity `label'.  This requires an increase in complexity
in the model, but allows us to include one of the most important galaxy
properties naturally in our models.
An alternative formulation would slice the 2D parameter space along the
other axis and specify $N(M|L)$ separately for bins of different $L$.  While
these are mathematically equivalent, it seems more natural within the halo
framework to specify the distribution of $L$ at fixed $M$ rather than the
reverse.  Guzik \& Seljak (\cite{GuzSel}) suggested an alternative way of
including luminosity information in the halo model, which we shall not pursue.

With the CLF, combined with the halo mass function $n(M)$, one can reconstruct
the luminosity function of galaxies:
\begin{equation}
  \Phi(L) = \int \Phi(L|M) {dn\over dM} dM
          = \bar{\rho}\int {\Phi(L|M) \over M}f(\nu)d\nu \;,
\label{eqn:one}
\end{equation}
where $dn/dM$ is the mass function and $f(\nu)$ is the multiplicity 
function (see Appendix \ref{sec:clfappendix}).
Also the (large-scale, linear) bias of galaxies as a function of luminosity
can be computed within this framework:
\begin{eqnarray}
  b(L) &=& {1\over \bar{n}} \int \Phi(L|M)b(M){dn\over dM}dM \;, \\
       &=& {\bar{\rho}\over \bar{n}} \int{\Phi(L|M)\over M} f(\nu)b(\nu)d\nu
  \qquad .
\end{eqnarray}

We follow Yang et al.~(\cite{YMvdB}) and model the conditional luminosity
function as a Schechter function,
\begin{equation}
  \Phi(L|M) dL
  = {\widetilde{\Phi}_* \over \widetilde{L}_*} {\left(L\over \widetilde{L}_*\right)}^{\widetilde{\alpha}} e^{-L/ \widetilde{L}_*} dL \;,
\end{equation}
with the three functions: $\widetilde{\alpha}$,
$\widetilde{L}_*$ and $\widetilde{\Phi}_*$.  Here the tilde
distinguishes these implicit 
functions of the halo mass, $M$, from the 
three parameters of the global luminosity function.
The full parameterization of CLF is derived based on
arguments about the total mass-to-light ratio of a halo and the
characteristic luminosity of a halo.
\begin{equation}
\left\langle M\over L\right\rangle(M) = {1\over 2} \left(M\over L\right)_0 
	\left[ \left(M\over M_1\right)^{-\gamma_1}+
	\left(M\over M_1\right)^{\gamma_2}\right] \;,
\end{equation}
\begin{equation}
{M\over \widetilde{L}_*(M) } = {1\over 2} \left(M\over L\right)_0 
	f(\widetilde{\alpha})\left[\left(M\over M_1\right)^{-\gamma_1}
	+\left(M\over M_2\right)^{\gamma_3}\right] \;.
\end{equation}
With the pre-factor,
\begin{equation}
f(\widetilde{\alpha}) = {\Gamma(\widetilde{\alpha} +2) \over
	\Gamma(\widetilde{\alpha}+1, 1)} \;,
\end{equation}
introduced to make $N_*(M) = 1$ for small mass halos ($M \ll \min[M_1,M_2]$). 
The mass-to-light ratio and the typical luminosity are both
assumed to have a broken power law form, arguments for which are given
in Yang, Mo \& van den Bosch~(\cite{YMvdB}).

For $\alpha(M)$, a simple linear function of $\log(M)$ is adopted,
\begin{equation}
  \widetilde{\alpha}(M) = \alpha_{15} + \eta \log_{10}(M_{15}) \;,
\end{equation}
with $M_{15}$ the halo mass in units of $10^{15}h^{-1}M_\odot$ and
$\alpha_{15} = \widetilde{\alpha}(M_{15}=1)$.

There is a minor problem caused by this definition of $\alpha(M)$ in
that the 
function $f(\alpha)$ becomes divergent when $\alpha\lesssim -1.3$.
For negative $\eta$ this happens at high mass, where
$\widetilde{L}_*(M)\sim M^{1-\gamma_3}/f(\alpha)$,
leading to an unphysical decrease in $\widetilde{L}_*$ with increasing mass.
The drop in $\widetilde{L}_*$ causes a drop in $N(M,L>L_*)$.
Fortunately this only happens for very high mass halos which are rare enough
that they do not affect the statistical properties which concern us here.

Because the Schechter form is assumed, it can be shown that
\begin{equation}
  \left\langle L\right\rangle(M)
  = \int L\Phi(L|M) dL
  = \widetilde{\Phi}_* \widetilde{L}_* \Gamma[\widetilde{\alpha}+2] \;,
\end{equation}
allowing $\widetilde{\Phi}_*$ to be derived from the above expressions.

There is an additional parameter $M_S$ which is the mass scale where the 
local $\widetilde{L}_*$ equals the global $L_*$, i.e. 
\begin{equation}
\widetilde{L}_*(M_S) = L_*
\end{equation}
Satisfying this requirement actually sets $(M/L)_0$. 

Thus in total there are 8 free parameters in the CLF algorithm:
$\alpha_{15}$, $\eta$, $M_1$, $M_2$, $\gamma_1$, $\gamma_2$,
$\gamma_3$, $M_S$. For a given set of parameters one can reconstruct the 
luminosity function and luminosity-dependent bias (semi-) analytically. 
Also the HOD can be computed from
\begin{equation}
  N(M,L>L_{\rm cut}) = \int \Phi(L|M) dL
              = \Phi_* \Gamma[\alpha+1,{L_{\rm cut}\over
              \widetilde{L}_*(M)}] \;.
\end{equation}
With the HOD and CLF, we are able to populate an N-body simulation with
galaxies and compare the luminosity function, bias and correlation function
with observations (see Appendix \ref{sec:nbody}). 

\section{Observational Constraints} \label{sec:constraints}

In this paper we wish to search for evolution in the way in which galaxies
populate halos.  We do this by constraining the conditional luminosity
function at $z=0$ and seeing if recent results from DEEP2 allow us to
rule out the `null hypothesis' that this CLF is independent of redshift.

Because the observations from these two surveys, and in
particular how they relate to the different selection effects
inherent in each survey, are so key to this undertaking we review here
in some detail the observational constraints we will adopt in our
subsequent analysis.

\subsection{The Galaxy Surveys}

The 2dFGRS (Colless et al.~\cite{2dF}) has recently been completed, having
observed redshifts for 230,000 galaxies over $\sim2000$ sq.~degrees on the
sky.  The galaxies targeted have been selected to
an extinction corrected $b_{\rm J}<19.5$ magnitude limit from an
updated version of the APM catalogue (Maddox et al.~\cite{maddoxa};
Maddox, Efstathiou \& Sutherland~\cite{maddoxb}).
As will be demonstrated, the sheer size of this data set allows us to
rigorously constrain our models at $z\sim0$ for later propagation to
higher redshifts.

The DEEP2 Galaxy Redshift Survey has recently begun to acquire large
numbers of galaxy redshifts, and preliminary analyses of the galaxy population
and its clustering behavior have recently been published
(Coil et al.~\cite{coil}; Madgwick et al.~\cite{mad03c}).  
This survey comprises objects selected from CFHT photometry that have
been pre-selected to have redshifts $z>0.7$ (Davis et al.~\cite{DEEP2}),
and is selected to be a magnitude-limited sample based upon the observed
$r_{\rm AB}$ magnitudes to $r<24.1$.

It is clear that, because of the different bands adopted for the
magnitude-based selection of the objects in these two surveys ($b_{\rm
J}$ in the 2dFGRS and $r_{\rm AB}$ in DEEP2), they may not in fact
probe identical galaxy populations.  For this reason we begin by
comparing the relative mix of different spectral types in these two
surveys before proceeding to describe in more detail the various
observational constraints from each.

\begin{figure}
\begin{center}
\resizebox{3.5in}{!}{\includegraphics{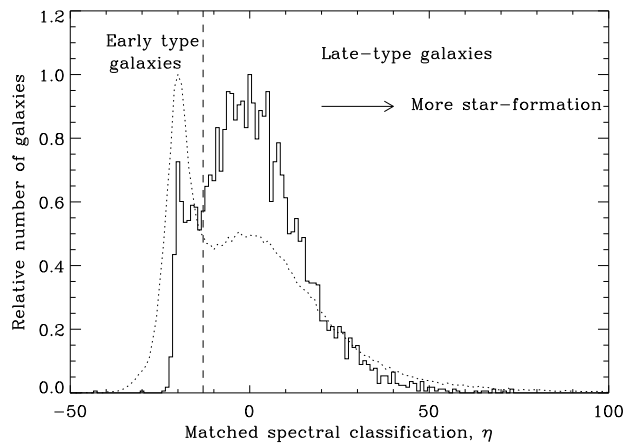}}
\end{center}
\caption{The distribution of different types of galaxies in the 2dFGRS
and DEEP2 are compared using a `matched' spectral classification (see
Madgwick et al.~\cite{mad03c} for details).  The dotted histogram
shows the distribution of galaxy types from the 2dFGRS and the solid
histogram shows that from the DEEP2 Survey.
From this comparison we can conclude that the DEEP2 Survey contains a
proportionately larger sample of so-called `late-type' galaxies.  For
this reason in what follows we compare the results from the DEEP2
Survey to those of the late-type 2dFGRS galaxies -- as well as for
the full 2dFGRS sample.}
\label{fig:competa}
\end{figure}

\subsubsection{Selection effects and spectral types}
\label{sec:selection}

To investigate the relative importance of selection effects in the two
surveys, 
we adopt here the spectral classifications developed for the 2dFGRS
and DEEP2 in 
Madgwick et al.~(\cite{DSMLF}) and Madgwick et al.~(\cite{mad03c})
respectively.   
These classifications, $\eta_{\rm 2dF}$ and $\eta_{\rm DEEP}$, have
been derived in an analogous way from 
a principal components analysis (PCA) of the galaxy spectra in each
survey and can hence be used to compare the relative mix of different
types of galaxies in each survey.
The classification itself
provides a continuous parameterization of the spectral type of a 
galaxy based upon the strength of nebular emission present in its 
rest-frame optical spectrum.   
It is found that $\eta$ correlates 
relatively well with galaxy $B$-band morphology at $z\sim0$
(Madgwick~\cite{mad03}).   
However, the most natural interpretation of 
$\eta$ is in terms of the relative amount of star formation 
occurring in each galaxy (Madgwick et al.~\cite{mad03a}).

In Fig.~\ref{fig:competa} we compare the distributions of spectral
types in the two surveys.  It is clear from this comparison that the
DEEP2 survey comprises relatively more galaxies that are undergoing
recent star-formation (given that it contains proportionately more
galaxies with high $\eta$ values).  This is in fact not surprising
given the different selection criteria in each survey.  In the case of
the 2dFGRS, the galaxies have been selected in the observed $b_{\rm J}$ 
band -- which gives preference to galaxies with recent star formation.
However, the observed frame $r$ selection adopted for the DEEP2
galaxies (which are typically $z\sim1$) in fact corresponds to a
rest-frame $U$-band selection -- which is even more biased towards
selecting galaxies with recent episodes of star formation.

It is clear from these arguments and Fig.~\ref{fig:competa} that
making a direct comparison between the 2dFGRS and DEEP2 galaxy
populations would not give a fair measure of evolution.
However those galaxies in the 2dFGRS with more recent star formation
may be a more analogous galaxy population.

In the analysis that follows we perform our fits for both the full
2dFGRS galaxy population (referred to as `all' galaxies) and for just
those galaxies with recent star formation (the `late-type' galaxies).
In so doing we can effectively bracket the galaxy population that is
actually observed in the DEEP2 Redshift Survey\footnote{Note that as more
data becomes available in the DEEP2 Survey, a much more detailed
analysis will become possible, in that we will be able to directly
compare observations based upon different types of galaxies in this
survey and the 2dFGRS.}.

\subsubsection{Galaxy clustering}

A key element of the analysis presented here is a detailed
characterization of the galaxy clustering at both $z=0$ and $z\sim1$.
At $z=0$ we have several observational constraints on the two-point
correlation function, $\xi(r)$, available to us (e.g. Norberg et
al.~\cite{Norberg01}; Norberg et al.~\cite{Norberg02}; 
Hawkins et al.~\cite{hawkins}; Madgwick et al.~\cite{mad03b}).
Because the actual 
type of galaxies under consideration appears to be so important to
making a fair comparison between the 2dFGRS and DEEP2, we choose to
adopt the $\xi(r)$ estimates from Madgwick et al.~(\cite{mad03b}),
since these 
have been performed for both `early' and `late-type' galaxies.

In terms of quantifying the degree of clustering at $z\sim1$ we have a
much more limited selection.  Coil et al.~(\cite{coil}) have recently
presented initial estimates for the correlation function in the DEEP2
Redshift Survey for a variety of different samples.
Because this survey occupies such a large range of redshift ($0.7<z<1.4$),
we choose to adopt the constraints that have been placed on the correlation
function over the more limited range $0.7<z<0.9$, so that we can neglect
evolution within this sample.
Specifically we shall assume that $\xi(r)$ is measured at the effective
redshift of this sample, as calculated by Coil et al.~(\cite{coil}),
$z_{\rm eff}=0.8$.
We also emphasize that the analysis presented by Coil et al.~(\cite{coil})
made use of only 2219 galaxies drawn from a contiguous sub-region of the
initial observations, and hence only represents $\sim5$\% of the total
potential of the survey.  We expect that constraints on the clustering of
galaxies at $z\sim 1$ will improve dramatically in the near future.

Note that preliminary estimates of the correlation function have also
been made for the types of galaxies in this survey analogous to the
2dFGRS, which would allow us to make much more detailed comparisons in
which selection effects would be less important.  However, given the
small size of the samples available at this early stage no meaningful
constraints can be obtained from these calculations.

\subsubsection{The luminosity function}

The LF provides a fundamental constraint in that it allows us to
restrict the number density of galaxies at different epochs.
We use here the galaxy LF estimates made by Madgwick et al.~(\cite{DSMLF}),
since these have been determined for different sub-samples of galaxy
types, which are applicable to the arguments of Sec.~\ref{sec:selection}.
This LF has been fit as a series of step functions (the SWML fit,
see Efstathiou, Ellis \& Peterson~\cite{eep}).

An important consideration in fitting to the LF is the fact that the
normalization, $\phi_*$, of the LF has been determined independently
of its shape.  The normalization of the LF therefore provides a
relatively independent constraint on $\phi(L|M)$ and will be treated as
such in our analysis.

\subsubsection{Absolute bias}

The absolute bias, $b(L_*)$, of galaxies in the 2dFGRS has been determined by
several authors over a range of different scales
(Verde et al.~\cite{verde}; Lahav et al.~\cite{lahav}; Hawkins et
al.~\cite{hawkins}).   
For our purposes we are most interested in the large-scale (linear) bias
regime -- which is most readily accessible to the analytic model used in
the HOD.
For this reason we do not adopt the bispectrum analysis presented in
Verde et al.~(\cite{verde}), which probes significantly non-linear scales.
The determination of $b$ made by Lahav et al.~\cite{lahav} and
Hawkins et al.~(\cite{hawkins}) are in the fully linear and {\em quasi}-linear
regimes respectively, with the Lahav et al. estimate 
corresponding to the largest scales. 
However, we choose to adopt the redshift-distortion based determination of
$b$ from Hawkins et al.~(\cite{hawkins}), 
since this has also been determined for
different types of galaxies (Madgwick et al.~\cite{mad03b}), 
allowing us to easily
extend our analysis to incorporate only the late-type galaxies we
expect to dominate in the DEEP2 Redshift Survey.

Given the numerous uncertainties involved in determining $\langle b\rangle$
we choose a conservative value $b(L_*)=1\pm 0.2$.  As it will turn out the
fit already prefers certain values of $b(L_*)$, our loose constraint will not
be too critical.

\subsubsection{Relative bias}

The observed bias is known to vary depending on several aspects of the
galaxy population under consideration.  Arguably one of the most
fundamental variations is the change in relative bias with the intrinsic 
luminosity of the galaxy population under consideration.  Norberg et
al.~(\cite{Norberg01}) measured the bias as a function of luminosity,
using the clustering of $L_*$ galaxies as a reference point.
They found the points can be fitted by a linear relation:
\begin{equation}
  b(L,z=0)/b(L_*,z=0) = 0.85 + 0.15(L/L_*)
  \qquad .
\end{equation}
Norberg et al.~(\cite{Norberg02}) found it is the luminosity,
not the type, that is the dominant factor causing the variation in the
clustering strength.
Also, Madgwick et al.~(\cite{mad03b}) show that, at large scale,
early and late type galaxies have almost the same clustering strength.
We will include the luminosity dependence of $b$, but neglect the type
dependence in the following.

\section{Parameter fitting} \label{sec:MCMC}

We have in total 9 free parameters that need to be constrained using
the results we have listed from the 2dFGRS.  For this reason a simple
grid-based search of the parameter space is not computationally
feasible.  Rather
we choose to adopt a Markov Chain Monte Carlo (MCMC) method to explore the
multi-dimensional space.  This has significant computational
advantages over a grid-based search and in particular it allows us to
more efficiently probe the maximum in the likelihood space.

For every random walk step used by the MCMC, we evaluate 
its $\chi^2$ using; 
\begin{enumerate}
\item the
observed luminosity function -- 22 normalized SWML points between
$0.05L_*$ and $5L_*$;
\item the relative bias points in 6 luminosity bins and a
rough estimate of the absolute bias $b(L_*)$;
\item the four priors we have added (\S\ref{sec:prior2}). 
\end{enumerate}
We will additionally incorporate constraints from the observed
correlation functions after the MCMC has been completed using an
`importance re-sampling' of the chain, as described later
(\S\ref{sec:important}).

Note that because the normalization and the shape of the luminosity
function provide relatively independent constraints, we want to treat
them separately. So, we fix the normalization of the analytically
computed points by the SWML luminosity bin nearest to $L_*$,
$\widehat{\Phi}(L_*)$, i.e.  
\begin{equation}
  \phi(L_i) = \Phi(L_i)
    {\widehat{\Phi}(L*)\over \Phi(L*)},
\end{equation} 
We then sum the $\chi^2$ of the remaining $N_l-1$ 
computed $\phi(L_i)$ points using SWML points and error
bars. This way we get the $\chi^2$ of the shape of the luminosity function. 
For the normalization, it's not determined as accurate as the shape. We set 
$\Phi(L_*)$ to have a fractional error of 10\% around the observed value 
$\widehat{\Phi}(L_*)$.

For the bias, since the observational results are given in relative bias
already, we do not need to do the same thing as we did for luminosity
function.  However, we do need to incorporate the $\chi^2$ of the
absolute bias.  

The $\chi^2$ is evaluated through the following equation.
\begin{equation}
  \chi^2 = \chi_{\Phi}^2+\chi_b^2 + \chi_{\rm priors}^2 \;,
\end{equation}
where the first two terms are;
\begin{equation}
  \chi^2_{\Phi} =
    {3\over N_l -1} \sum_{i=1}^{N_l} \left[\widehat{\Phi}(L_i) -\phi(L_i)
    \over \Delta\widehat{\Phi}(L_i)\right]^2 
    + \left[\widehat{\Phi}(L_*) - \Phi(L_*) \over 0.1
    \widehat{\Phi}(L_*)\right]^2 \;,
\end{equation}
and
\begin{eqnarray}
\chi^2_b = {1\over N_b}\sum_{i=1}^{N_b}
    \left[{\widehat{b}(L_i)-b(L_i)\over\Delta \widehat{b}(L_i)}\right]^2 
  +{\left[b(L_*)-1\over 0.2\right]}^2 \;.
\end{eqnarray}
Here $\widehat{\Phi}(L_i)$ and $\widehat{b}$ are
observed quantities and all symbols without a `hat' are analytically
computed quantities. 
We give the shape of the luminosity function a relative weight 3 since
it is measured with much less uncertainty than other constraints we used.
As it turns out the constraint $b(L_*)=1\pm 0.2$ is not really needed as
the fit will provide a tighter constraint.

The chain is started from the model used in
Yang, Mo \& van den Bosch (\cite{YMvdB}) and is then allowed a
`burn-in' period for the chain to equilibrate in the
likelihood space.
The cosmology used in the code is the same as in the simulation.
At any point in the chain we generate a new trial element by drawing
parameter shifts from independent Gaussian distributions in each of the
nine CLF parameters.
The probability of accepting a new random walk step is taken to be
\begin{equation}
  P_{\rm accept} =\left\{
    \begin{array}{cc} 
    1.0 & \mbox{for $\chi^2_{\rm new} < \chi^2_{\rm old}$} \\
    \exp\left[-(\chi^2_{\rm new} -\chi^2_{\rm old})/2\right]
    & \mbox{for $\chi^2_{\rm new} > \chi^2_{\rm old}$} 
    \end{array} \right.
\end{equation}
Given the sampling strategy we adopted, the acceptance rate is around
25\%, leading to nearly independent samples in the chain after a
(few) thousand elements.
We ran four chains, with different random number seeds, each for 100,000
steps. 
The $\chi^2$ does not decrease very much from the initial guess, indicating
that the starting model is within the equilibrium region of the chain.

\subsection{Priors}
\label{sec:prior}
\label{sec:prior2}

The CLF formalism has a `free' function for halos of every mass.  Even given
the restriction to Schechter forms this requires us to choose three free
functions of mass ($\widetilde{\alpha}$, $\widetilde{L}_*$ and
$\widetilde{\Phi}$) to specify the model, and further parameterizing these
functions as (double) power laws still allows a great deal of freedom.

Unfortunately, we have found that it is possible to obtain reasonable
$\Phi(L)$ and $b(L)$ 
for `pathological' $N(M,L>L_{\rm cut})$.  For example, the number of galaxies
brighter than the local $\widetilde{L}_*(M)$ in a halo of mass M is
proportional to $M^{\gamma_3-\gamma_2}$.  If $\gamma_2$ becomes larger than
$\gamma_3$ and $\widetilde{L}_*(M)$ does not increase fast enough to
compensate, the number of galaxies brighter than (the global) $L_*$ in a
halo of mass $M$ will decrease for $M \gg \max[M_1,M_2]$.

To eliminate these and other instabilities we have chosen to apply some
priors on $N(M)$.
In most cases the pathological models would be rejected by comparing the
predicted and observed correlation functions (see \S\ref{sec:important}),
however we include those constraints after the chain has been run.
In this sense one can consider our priors as a way to increase the
efficiency of our procedure.

After extensive experimentation we found that the following additional
priors worked quite well in our analysis.  First we apply a Gaussian
prior that $\alpha_{15}=-1.0\pm 0.2$ to ensure that only negative values
of $\alpha_{15}$ are adopted by the chain.
Without this prior positive values of $\alpha_{15}$ occur quite regularly,
despite being unphysical.
In particular, we have based this prior upon composite observations of the
Coma cluster LF by Driver \& De Propris~(\cite{depropris}), who have derived
a faint-end slope of $\alpha=-1.01^{+0.04}_{-0.05}$, for this cluster.
We allow more flexibility in our constraint than this, to allow for the
fact that this is only a rough constraint on our model.

We also provide two priors on the shape of $N(M)$, both log-normal.
The first sets $\widehat{N}(10^{15}\,h^{-1}M_\odot)$ to a fiducial value
with a factor of 5 ($1\sigma$) error and the second provides a similar
constraint on $\widehat{N}(10^{12}\,h^{-1}M_\odot)$.
After the importance resampling described in \S\ref{sec:important} the
distributions of $\widehat{N}_{15}$ and $\widehat{N}_{12}$ are both much
narrower than the imposed priors, indicating that the exact form is not
important.

Our last prior provides a constraint on the number of galaxies residing in
halos of very low mass.  This prior has been included for both theoretical
and practical reasons and again the distribution in the chain after importance
resampling (\S\ref{sec:important}) is much narrower than the prior.
Theoretically we believe that even low luminosity galaxies do not live in very
low mass halos, $M\ll 10^{10}\,h^{-1}M_\odot$, and we wish to encode this
information in the fit.  The practical reason is that our N-body simulation
has limited mass resolution which makes it impossible for us to compute
$\xi(r)$ for such models.
Mathematically the problematic models arise when $L_*$ does not decline fast
enough with mass.  In this situation many of the fainter galaxies can
reside in halos of mass smaller than $10^{10}\,h^{-1}M_\odot$.
Even though $N(M,>0.1L_*)$ is in the range $10^{-4}$ to $10^{-3}$ for these
halos, their huge number ensures a significant contribution to the total
number of galaxies.
We require $L_*(M=10^{10}\,h^{-1}M_\odot)<0.02L_*$ so that almost no
$0.1L_*$ galaxies live in halos with $M<10^{10}\,h^{-1}M_\odot$.

\subsection{Importance re-sampling}
\label{sec:important}

The MCMC calculated so far has adopted only the constraints from the
2dFGRS LF and biasing results to constrain the CLF at $z\sim0$.  
In fact there is substantially more information available to us in the
form of very accurate determinations of the two-point correlation function,
$\xi(r)$.

It is possible to directly relate the CLF to the correlation function
analytically, by making various assumptions about the way halos and
galaxies are 
distributed relative to each other (e.g.~Peacock \& Smith~\cite{PeaSmi};
Seljak \cite{Sel00}; Maggliochetti \& Porciani \cite{MagPor};
Zehavi et al.~\cite{Zehavi03}).
However, given that we have detailed N-body simulations available to us,
we instead choose to populate these using the CLF and then calculate
$\xi(r)$ estimates from these.  

Using the N-body simulations instead of
analytic approximations to the halo profiles and clustering allows us
to make much more detailed comparisons to the shape of the correlation
functions, which is particularly appealing given the very high
fidelity of these observations.

One immediate drawback of calculating the correlation function from the
N-body simulations (instead of purely analytically), is the additional
computational time required.  Given that the MCMC has yielded a chain
of 400,000 elements, the calculation of $\xi(r)$ for each of these
is infeasible.
Fortunately it is also unnecessary, as the distribution of $\xi(r)$ can
be obtained from a smaller set of samples.
We choose 200 models, 
(given that only one in every few thousand
elements is completely independent),  at random, 
from the chain and compute $\xi(r)$ for
each of these, for the `all galaxy' and `late-type' samples at $z=0.1$
and $z=0.8$ (the 
effective redshifts of the 2dFGRS and DEEP2 $\xi(r)$ calculations).
We then adopt an `importance re-sampling' technique (Gilks, Richarson
\& Spiegelhalter \cite{gilk}) to include the
goodness-of-fit between the models and the observed 2dFGRS correlation
function.
We only fit $\xi(r)$ over the range $0.2<r<3\;h^{-1}$Mpc, since for
separations $r<200\;h^{-1}$kpc fibre collisions lead to a decrease in
the observed 2dFGRS $\xi(r)$ (see e.g.~Hawkins et al.~\cite{hawkins}),
and for $r>3\;h^{-1}$Mpc our N-body simulations become susceptible to
finite box size effects.

\subsection{Results}

\begin{figure*}
\begin{center}
\resizebox{3.5in}{!}{\includegraphics{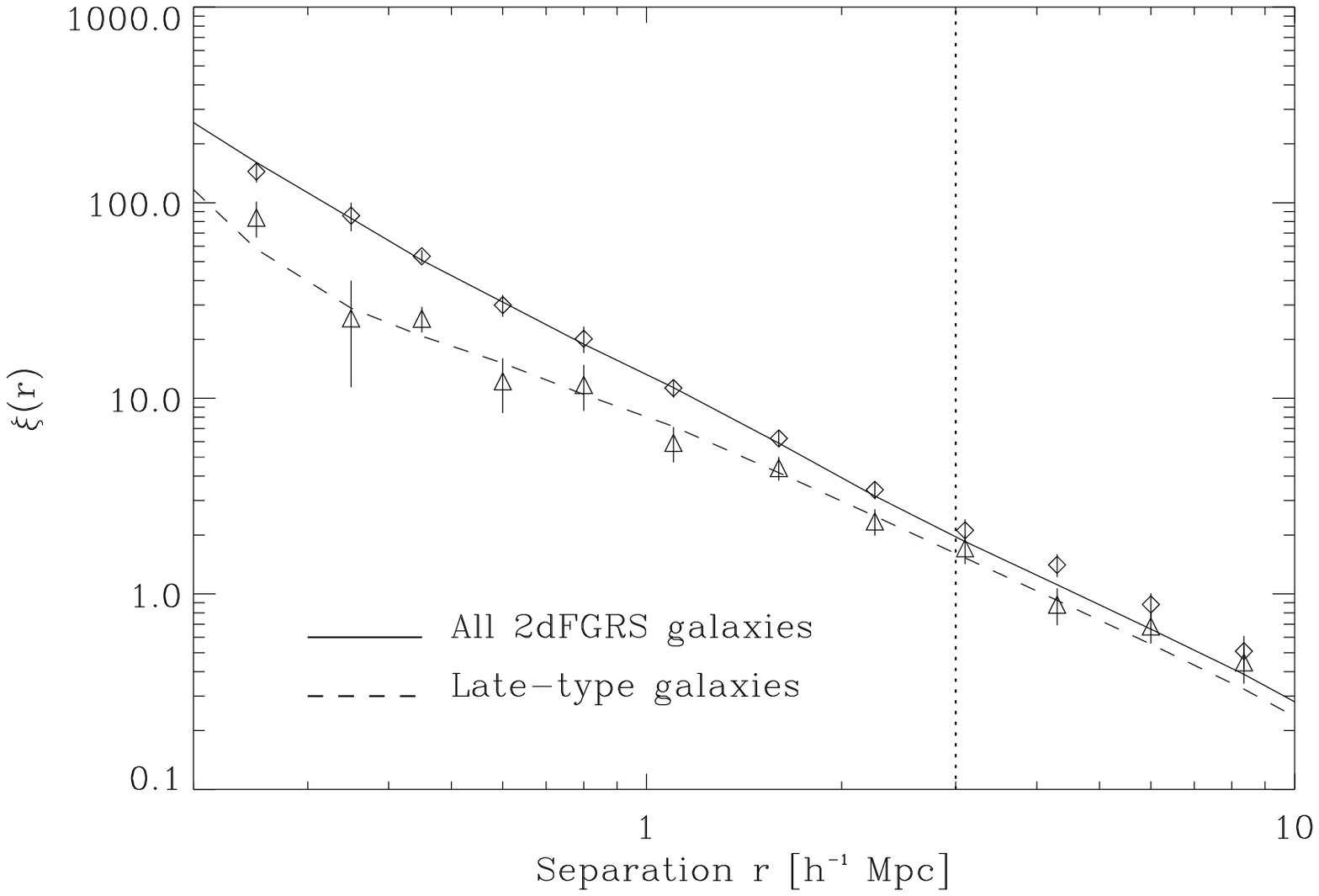}}
\resizebox{3.5in}{!}{\includegraphics{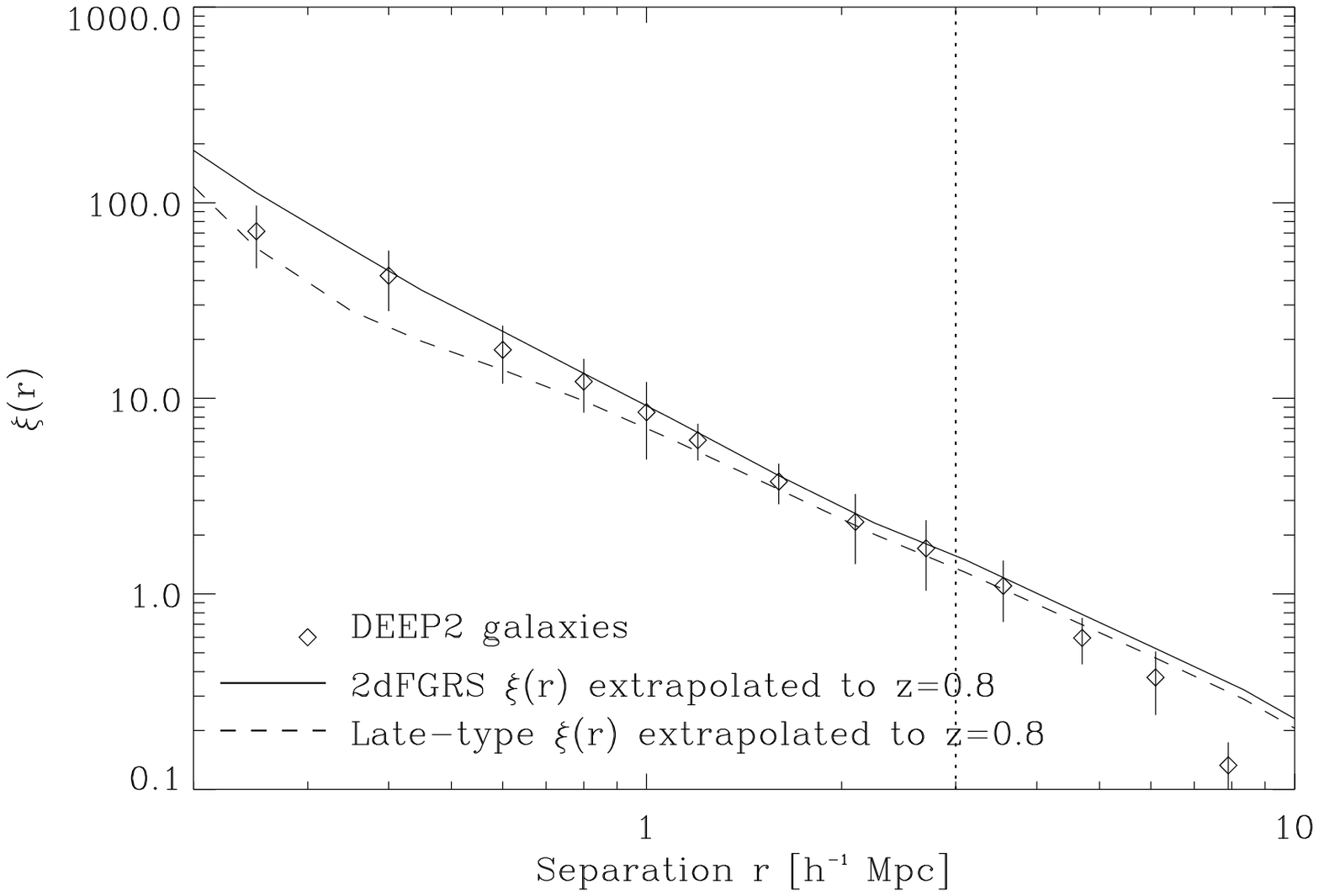}}
\caption{The left panel shows the mean correlation functions after
performing importance re-sampling on the model CLFs
fit to the 2dFGRS for `late-type' (dashed line) and
`all' (solid line) galaxies.
It can be seen that our models can recover the observed
$\xi(r)$ in great detail.  In each plot, the dotted vertical line at
$r=3\;h^{-1}$ Mpc shows the maximum separation over which our $\xi(r)$
estimates from the N-body simulations are robust.  The right panel shows the
correlation function calculated using the DEEP2 Redshift Survey
(points), and compares this with the mean correlation functions,
derived from the same model CLFs used in the left panel
(after extrapolating the N-body simulations to $z=0.8$).  The agreement
between the mean model CLF $\xi(r)$'s at $z=0.8$ and the observed DEEP2
$\xi(r)$ is very good and quite unexpected.  This result suggests
that we cannot rule out the null-hypothesis of no evolution in the
halo model between redshifts $z=0.1$ and $z=0.8$.}
\label{fig:xi}
\end{center}
\end{figure*}

Once the importance re-sampling is incorporated then all the necessary
$z\sim0$ observational constraints have been included in our CLF
determination and the parameter space can be reviewed.
In particular, we compare in the left panel of Fig.~\ref{fig:xi} the
{\em mean\/} correlation functions derived from our fitted CLFs and N-body
simulations to their corresponding observed $\xi(r)$ from the 2dFGRS.
It can be seen from this figure that the models have very successfully fitted
both the correlation function of the `late-type' and `all galaxy' 2dFGRS
samples. 

\begin{figure*}
\begin{center}
\resizebox{3.5in}{!}{\includegraphics{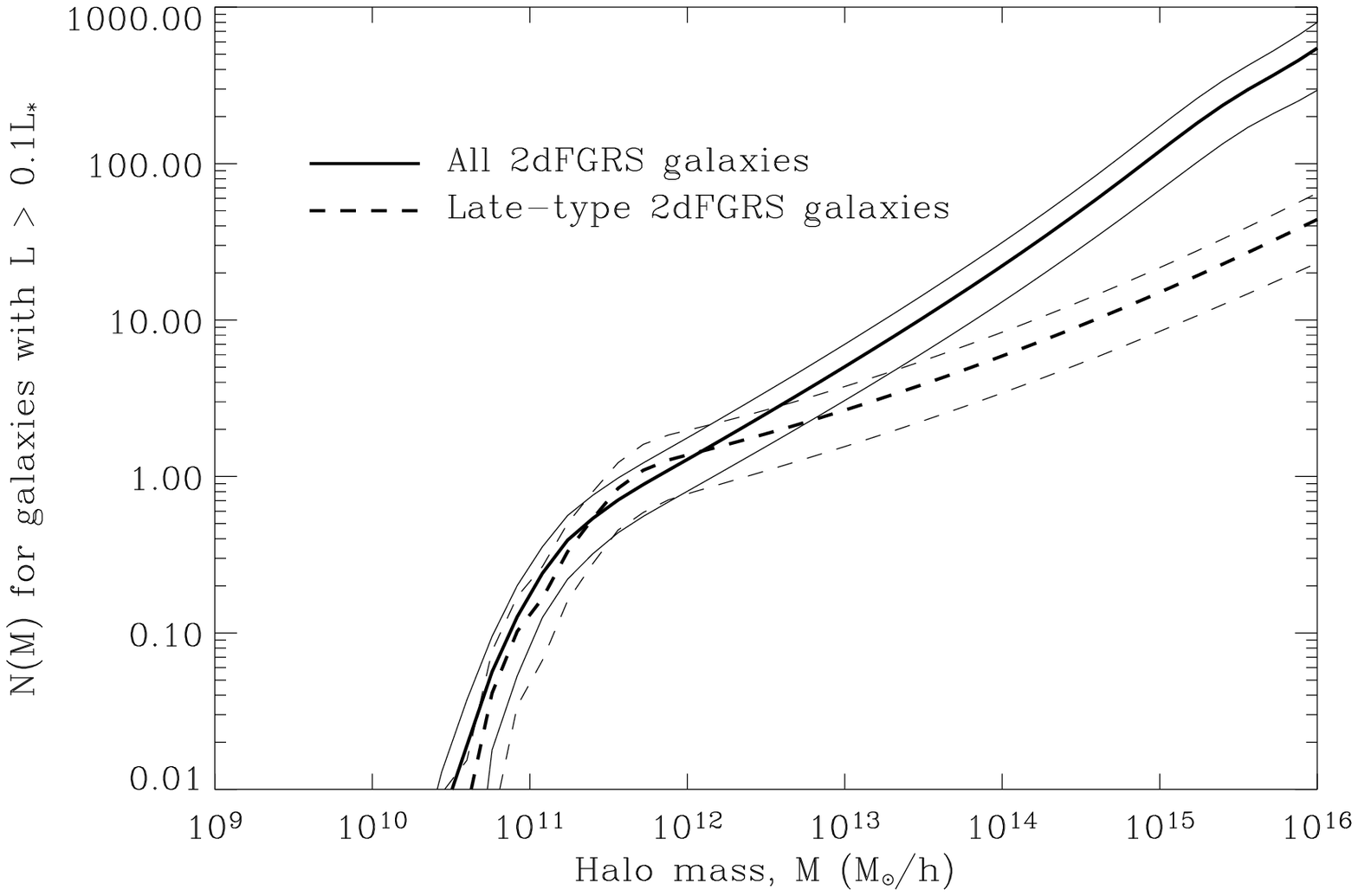}}
\resizebox{3.5in}{!}{\includegraphics{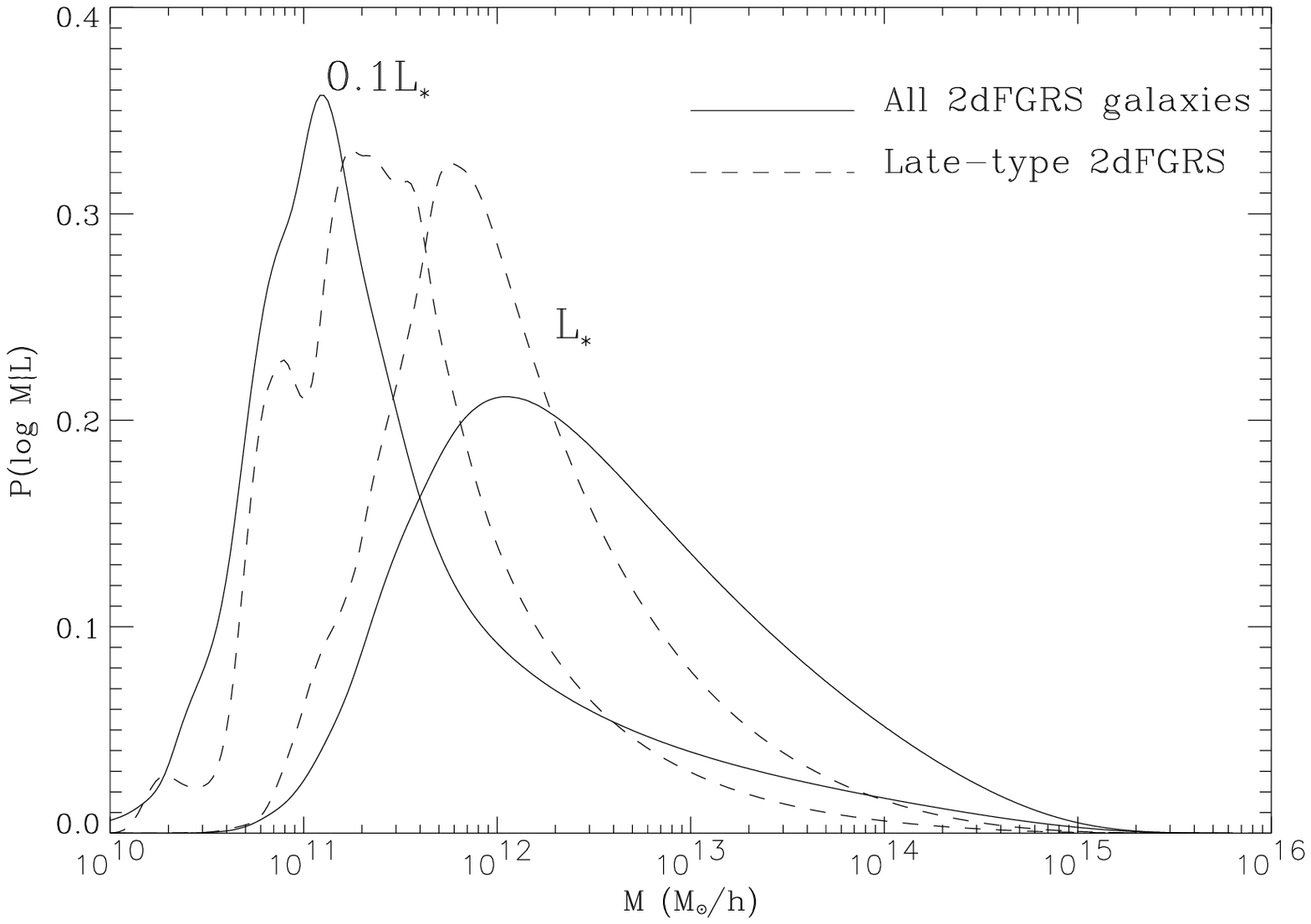}}
\end{center}
\caption{The mean HOD, $N(M)$, is shown for galaxies with $L>0.1L_*$
(thick lines; left panel),  as derived from the CLFs after importance
re-sampling to fit the observed 2dFGRS correlation function.  The variance
at each mass is indicated by the thinner lines bracketing the central one.
It can be seen that the late-type galaxies appear to dominate low mass
halos, but become steadily more under-represented in higher mass halos.
The right panel shows how we can invert this to derive the probability
that a galaxy with a given range of luminosity inhabits a halo of mass,
$M$.  This function, $P(M|L)$, is shown here for both $0.1L_*$ and $L_*$
galaxies, the solid lines again correspond to the `all galaxies' sample
and the dashed lines to the `late-type' sample.}
\label{fig:nm}
\end{figure*}

The mean HODs derived from our model CLFs are shown in Fig.~\ref{fig:nm}.
In this particular plot we show the mean number of $L>0.1L_*$ galaxies
occupying a halo of the specified mass.  This particular luminosity cut
is the one we have adopted throughout this analysis, as it is found to
correspond well to the observed distributions of luminosities in the
two surveys.

It can be seen from Fig.~\ref{fig:nm} that the two CLFs
predict very similar low-mass cut-offs in the HOD, but have very
different behavior at high masses.  Clearly star forming galaxies
appear to dominate the galaxies within low mass halos, but become
successively more
under-represented in the highest mass halos.  This result is certainly
not surprising given observed morphology-density relations
(e.g. Dressler~\cite{dres}), however it is interesting to see the
effect quantified in this way. 

A different way of plotting the CLF is to instead relate the
probability that a galaxy within a specified range of luminosities
inhabits a dark matter halo of a given mass.  This is also shown in
Fig.~\ref{fig:nm}, for both the `all galaxy' and `late-type' samples.
A similar plot appeared in Yang et al.~(\cite{YMvdBC}) for a single
representative model, our procedure allows us to also include the
uncertainty in the model parameters in a statistically correct manner.

\section{Constraining galaxy evolution} \label{sec:results}

Our preliminary analysis is complete, in that we have successfully
constrained the CLF formulation of the halo-model to the observations
of the 2dFGRS.  In so doing we have had to make only the most basic
prior assumptions suggesting that the observational data is more than
adequate to tightly constrain this model.

The next step in our analysis is to determine the degree of inference we
can make as to the evolution in the galaxy population to redshift
$z\sim0.8$, by contrasting the predictions of this constrained model
to the preliminary results from the DEEP2 Redshift Survey.

\subsection{Reconstructing the DEEP2 $\xi(r)$}

As discussed previously, the DEEP2 correlation function, $\xi(r)$, has
been determined in Coil et al.~(\cite{coil}), and it is this estimate
that we now try to recover from our CLF models.  In particular, we use
the $\xi(r)$ estimated over the redshift interval $0.7<z<0.9$ which
has an effective redshift $z_{\rm eff}=0.8$.

The simplest assumption we can make in propagating our model to higher
redshifts, is that the CLF has not changed between $z=0.1$ and $z=0.8$.
In order to test this assumption we again turn to our N-body simulations,
which allow us to extrapolate the dark matter halo distribution to $z=0.8$.
these simulations can then be re-populated and the correlation functions
recalculated.

The results of this analysis are presented in the right panel of
Fig.~\ref{fig:xi}, where 
the mean N-body $\xi(r)$ is shown for both the `late-type' and `all
galaxy' CLFs (after importance re-sampling with respect to the 2dFGRS
correlation functions at $z=0.1$), propagated to redshift
$z=0.8$. The fractional uncertainty is 20\%, independent of scale. 
What is remarkable from this
figure is the fact that the predicted correlation function at
$z=0.8$ from our fitted CLFs is completely consistent with
that which has been 
observed in the DEEP2 Survey -- especially for the late-type
2dFGRS galaxy sample which probably corresponds more closely to
the galaxies observed in the DEEP2 Redshift Survey (see
\S\ref{sec:selection}).
The agreement could be even better than shown in Fig.~\ref{fig:xi} if
we account for the effects of mask making on the DEEP2 data points at
small scales ($<300\,h^{-1}$kpc) and finite field effects at larger
scales.

This result is extremely surprising, in that it appears to imply that
we cannot rule out an unevolving halo model to $z\sim1$: the galaxies
observed in the DEEP2 Redshift Survey appear to inhabit their parent
dark matter halos in the same way as those observed in the 2dFGRS. 
Had the cut-off in $N(M,>0.1L_*)$ evolved to higher masses we would have
seen a steepening in $\xi(r)$ on sub-Mpc scales.  Also, had the $N<1$ tail of
$N(M,>0.1L_*)$ extended over a broader mass range at $z\sim 1$ than
$z\sim 0$ we would have seen a suppression of $\xi(r)$ on sub-Mpc scales.
Similarly, steepening the high-$M$ slope of $N(M)$ increases the bias,
which is not observed.

It is difficult to place precise limits on what functional forms are
allowed because of the limited amount of data, the uncertainty in the
underlying cosmological model and the large dimensional space in which
we are working (with its associated degeneracies).  As a rough gauge of
sensitivity, increases in the mass cut-off by even an order of magnitude
are strongly disfavored as is steepening $N(M)$ to $N\propto M$.  We
defer a more thorough analysis until more data become available.

\subsection{Other predictions for the DEEP2 Survey}

Given that the CLF does not appear to evolve significantly at high
redshifts, it is now possible to make a series of predictions relating
to what we expect to see in the DEEP2 Survey as more data becomes
available.  We briefly describe a handful of these predictions in this
section. 

\subsubsection{The luminosity function}

The evolution of the LF is perhaps one of the most natural outputs of
the conditional luminosity function.  Given that we can accurately
trace the dark matter halo mass function through N-body simulations,
Eqn.~\ref{eqn:one} provides a simple link between the two.

We have calculated the expected LF at $z=0.8$, given the constraints
placed on the CLF from the 2dFGRS observations. 
Interestingly it shows little evolution -- only a small variation 
in the faint-end slope, $\alpha$, and the characteristic luminosity,
$L_*$, are seen to $z=0.8$. 
This is an interesting result which suggests that our hypothesis
of a non-evolving CLF also implies little evolution in the {\em relative\/}
distribution of luminosities of galaxies.
Note that our clustering analysis has been determined entirely in terms of
relative luminosities ($L/L_*$) so evolution in the absolute value of $L_*$
is not constrained.
In fact, given that we expect the distribution of galaxies to be relatively
established by $z=0.8$ (as would be expected for no evolution in the CLF),
a purely {\em passive\/} change in $L_*$ would be entirely consistent with our
results.  Such a change would be brought about by e.g.~the internal dimming
of the galaxies as their stellar populations became successively older
and fainter.
However, our prediction of the distribution of luminosities relative to
$L_*$ should be robust, and so for this reason we predict that little
evolution in the faint end slope of the LF will be seen in the DEEP2 Survey,
when compared to the 2dFGRS.
Once a detailed calculation of the DEEP2 LF becomes available we will
pursue this possibility in more detail.

\subsubsection{Luminosity dependent bias}

The average bias at $z\sim 0$ has a luminosity dependence close to
that reported by Norberg et al.~(\cite{Norberg01}), since these data
were used in the fit.  Overall we find that the bias is determined
by the data to 2\% for $L\ll L_*$ and 5\% for $L\gg L_*$, both much
stronger constraints than our prior on $b$.
We can then ask how the mean bias relation evolves with redshift,
again assuming that the CLF is independent of $z$.
By $z=0.8$ the mean bias has risen 30\%, and the variance across the
chain is between 5-9\% depending on $L$.  The luminosity dependence
has steepened from $0.15$ to $\sim 0.25$ as shown in
Table \ref{tab:bias}.

\begin{table}
\begin{center}
\begin{tabular}{c|c|c}
$L/L_*$ & $b(z=0.1)$      & $b(z=0.8)$ \\ \hline
  0.3   & $0.94\pm 0.03$ & $1.13\pm 0.05$ \\
  0.7   & $1.02\pm 0.03$ & $1.26\pm 0.05$ \\
  1.0   & $1.08\pm 0.04$ & $1.34\pm 0.06$ \\
  1.5   & $1.18\pm 0.05$ & $1.48\pm 0.09$ \\
\end{tabular}
\end{center}
\caption{The large-scale, linear theory bias at $z=0$ and $z=0.8$ from
our importance resampled chains for all galaxies.  The mean and variance
defined by the 200 elements in the chain are reported.}
\label{tab:bias}
\end{table}

\section{Conclusions} \label{sec:conclusions}

In this paper we have attempted to constrain the degree of evolution
in the way galaxies inhabit their parent dark matter halos by
contrasting the very accurate observations of the galaxy population 
in the 2dFGRS (at $z\sim 0$) to
preliminary results from the DEEP2 Redshift Survey at $z=0.8$.
In order to do so we have had to carefully consider the fidelity of
the different observational results available for this analysis and
how these relate to the different selection effects in each survey.

Based upon a comparison between the types of galaxies observed in each
survey, we were able to conclude that the DEEP2 Redshift Survey
contains a much larger fraction of `late-type' galaxies i.e. those
currently under-going 
significant amounts of star formation.  
For this reason we have attempted to not only contrast
this survey with the full 2dFGRS, but also with observational results
based upon only the `late-type', star forming galaxies in this survey.
In this way we believe we are able to make a much more fair comparison
between the two surveys, in that we can effectively `bracket' the
population observed in the DEEP2 Redshift Survey.  Note that as more
data becomes available in the DEEP2 Survey, this bracketing will no
longer be necessary as we will be able to directly compare the
different types of galaxies in each survey.

A great deal more observational results are available at $z=0$ than at
$z\sim1$, and for this reason we have decided to adopt the approach of
making the most detailed fit possible of the HOD at $z=0$, and to then
test how consistent this HOD was to preliminary clustering results at
$z=0.8$.  In so doing we were able to show that the current
observation of the correlation function at $z=0.8$ is remarkably
consistent with no evolution in the HOD to this redshift.

This result is very surprising.  The HOD quantifies the impact of
galaxy formation and interaction processes present in the galaxy
population, and in particular how this relates to the underlying
dark-matter halo distribution.  Our results suggest that there has in
fact been no change in the way galaxies are distributed in their host
dark-matter halos over approximately half the age of the Universe.

At present our results are preliminary, in that the correlation
function of galaxies at $z=0.8$ is still relatively poorly
constrained.  In addition, the absence of an estimated luminosity
function for this population makes a more detailed comparison of the
evolutionary processes difficult.  However, as the DEEP2 Redshift
Survey continues to increase its sample size we expect these issues to
be resolved so that much more conclusive remarks can be made.  In
addition, as the sample size increases, correlation functions for
different types of galaxies in the DEEP2 Redshift Survey will become
available, which will allow us to be much more precise about the
importance of galaxy type selection in our results.

Other
improvements to our analysis will be forthcoming when results from the
recently begun
VLT-VIRMOS Redshift Survey (Le Fevre et al.~\cite{lefevre}) 
become available.  This survey comprises an
$I_{\rm AB}$-selected sampling of the galaxy population at similar
redshifts to the DEEP2 Redshift Survey, which at these redshifts is
much more comparable to a rest-frame $B$-selection of galaxies.  For
this reason the incorporation of this survey into our analysis should
allow us to more accurately address the issues of sample selection and
to what degree these impact our conclusions.

\acknowledgements

We would like to thank Marc Davis, Brian Gerke and Joanne Cohn 
for helpful discussions, and Alison Coil for providing the DEEP2
correlation function in electronic format.  
M.W.~would like to thank Ravi Sheth and
Andreas Berlind, and D.S.M.~thanks Shaun Cole
for numerous enlightening discussions on the halo
model.
The simulations used here were performed on the IBM-SP at the National
Energy Research Scientific Computing Center.
This research was supported by the NSF and NASA.

\appendix

\section{The Halo Model} \label{sec:clfappendix}

In this appendix we describe the details of the halo model formalism that
we use in the main paper.  A review of the halo model, with reference to
the original literature, can be found in Cooray \& Sheth (\cite{CooShe}).

The halo mass function, $dn/dM$, describes the (comoving) number density
of halos with mass in the interval $[M,M+dM]$.
This function depends on the redshift and the power spectrum of
dark matter halos as well as the underlying cosmology.  Typically one
adopts a change of variables, to 
give a dimensionless form of this function in terms of the halo peak height
$\nu$.
Specifically, the multiplicity function, $f(\nu)$, is used
since this is independent of cosmology.  The relation between this
multiplicity function and the halo mass function is simply given by,
\begin{equation}
  {\bar{\rho}\over M}f(\nu) d\nu = {dn\over dM} dM \;.
\end{equation}
where $\bar{\rho}$ is the mean matter density of the Universe, and the
cosmology dependence resides entirely in the peak height definition,
\begin{equation}
  \nu \equiv [\delta_c(z)/\sigma(M)]^2 \;.
\end{equation}
Note that many authors use $\nu=\delta_c(z)/\sigma(M)$ rather than its
square as we have done here.
Here $\delta_c \simeq 1.868$ is a threshold parameter taken from the theory
of spherical top-hat collapse while $\sigma(M)$ is the rms mass fluctuation
within spheres of radius $R=[ M/(4\pi\bar{\rho}/3)]^{1/3}$ evaluated in
linear theory, at redshift $z$.
For the multiplicity function, $f(\nu)$, we use the `ST' form, motivated by
ellipsoidal collapse and fit to N-body simulations
(Sheth \& Tormen ~\cite{SheTor}):
\begin{equation}
  \nu f(\nu) = A (1+\nu^{\prime-p})\nu^{\prime 1/2}e^{-\nu^\prime/2} \;,
\end{equation}
where $p=0.3$ and $\nu'=0.707\nu$.
The constant $A$ determines the normalization and is fixed by the requirement
that all the mass lie in a given halo,
\begin{equation}
  {1\over \bar{\rho}}\int M {dn\over dM}dM = \int f(\nu) d\nu = 1 \;.
\end{equation}
We can regain the older Press \& Schechter~(\cite{PS}) 
expression by taking $\nu'=\nu$ and $p=0$.

Halos within some mass range $[M,M+dM]$ are biased tracers of the
underlying matter distribution.  To linear order the bias can be
computed from the peak-background split
(Efstathiou et al.~\cite{EFWD}; Cole \& Kaiser \cite{ColKai};
Mo \& White \cite{MoWhi})
which for the ST mass function gives,
\begin{equation}
  b(\nu) = 1+ {\nu^\prime-1 \over \delta_c} + {2p\over \delta_c
           (1+\nu^{\prime p})} \;.
\end{equation}
This bias is appropriate for very large scales where the bias is
expected to be deterministic, linear and scale independent.
Note that this scheme automatically satisfies the requirement that the
mean bias of mass is unity:
\begin{equation}
  \left<b\right>_{\rm mass} = \int b(\nu)f(\nu)d\nu =1  \qquad .
\end{equation}

Given a HOD function $\left\langle N\right\rangle(M)$ we can use the mass
function and halo biasing scheme
described above to easily compute the mean galaxy bias (which is
defined as the scale independent linear bias of galaxy relative to mass),
\begin{eqnarray}
  \left\langle b\right\rangle_g =
  {\bar{\rho}\over \bar{n}} \int {\left\langle N\right\rangle\over M}
  f(\nu) b(\nu) d\nu \;.
\end{eqnarray}
Note that $\bar{n}$ is the mean density of galaxies, and can be computed
from,
\begin{eqnarray}
  \bar{n} = \int \left\langle N\right\rangle {dn\over dM} dM
          = \bar{\rho} \int {\left\langle N\right\rangle\over M}
            f(\nu) d\nu \;.
\end{eqnarray}

\section{Galaxy catalogues} \label{sec:nbody}

To compute the correlation function given the CLF we make use of a large
N-body simulation of a $\Lambda$CDM model.  The simulation employs
$512^3$ particles in a periodic, cubic box of side $128\,h^{-1}$Mpc
and was run with the {\sl TreePM\/} code described in White (\cite{TreePM}).
The cosmological model was chosen to provide a reasonable fit to a wide range
of observations with $\Omega_{\rm m}=0.3$, $\Omega_\Lambda=0.7$,
$H_0=100\,h\,{\rm km}\,{\rm s}^{-1}{\rm Mpc}^{-1}$ with $h=0.7$,
$\Omega_{\rm B}h^2=0.02$, $n=0.95$ and $\sigma_8=0.9$
(close to the best fit to the CMB data for this cosmology).
The gravitational force softening was of a spline form, with a
``Plummer-equivalent'' softening length of $9\,h^{-1}$kpc comoving.
The particle mass is $1.3\times 10^{9}\,h^{-1}M_\odot$.
The simulation was started at $z=100$ and evolved to the present with the
full phase space distribution dumped every $128h^{-1}$Mpc from $z\simeq 3$
to $z=0$.
As a check on finite volume effects we compared our results to those from
a similar simulation in a box with force softening and box side twice as
large, with $8\times$ more massive particles.  It is based on this comparison
that we restrict our fits to $r<3\,h^{-1}$Mpc.

We use outputs at redshifts appropriate to the 2dF ($z\simeq 0.1$)
and DEEP2 ($z\simeq 0.8$) surveys.
For each output we produce a halo catalogue by running a
``friends-of-friends'' group finder (e.g.~Davis et al.~\cite{DEFW}) with
a linking length $b=0.15$ (in units of the mean inter-particle spacing).
This procedure partitions the particles into equivalence classes, by linking
together all particle pairs separated by less than a distance $b$.
We keep all halos with more than 8 particles, and consider each of these
halos as a candidate for hosting `galaxies'.

We populate the simulation with `galaxies' by marking certain simulation
particles and assigning them luminosities.
The HOD function computed from the CLF model gives the mean number of
galaxies (more luminous than some $L_{\rm cut}$) which would be in a halo
of mass $M$.
The halo mass is estimated as the sum of the masses of the particles in the
FoF halo, times a small correction factor which provides the best fit to the
Sheth-Tormen mass function (Sheth \& Tormen \cite{SheTor}).
An actual number of galaxies is drawn from a distribution for each halo in
the simulation, and we use the nearby integer distribution.

Once the number of galaxies in each halo is known, they are assigned
luminosities from $\Phi(L|M)$.  The most luminous galaxy is assigned to the
center of mass of the halo, and the other galaxies are assigned to random
particles within the halo.  
While the code allows the possibility of a radial or velocity bias in
assigning galaxies to particles, throughout we assumed that galaxies traced
the mass and velocity distribution of the halo
(inheriting its shape and any substructure).

Given the total number of galaxies brighter than $L_{\rm cut}$ in each halo
it is necessary to choose luminosities for them based on $\Phi(L|M)$.
Just as it was necessary to specify both $\langle N\rangle$ and the higher
moments above, it is necessary to know the fluctuations about the mean
$\Phi(L|M)$ at this stage.  If the luminosities of galaxies each halo are
drawn independently from $\Phi(L|M)$ then one occasionally finds relatively
low mass halos with two `bright' galaxies.  Since such systems have small
radii this in turns implies an increase in the `bright' galaxy correlation
function at small scales.  Such an increase can indeed be seen in some of
the semi-analytic models, but appears to be absent in data.
This suggests that some mechanism acts to suppress pairs of bright galaxies
in small halos.  We can model this in a number of ways.  On one extreme we
could calculate the luminosities for all galaxies in halos of similar masses
by drawing from $\Phi(L|M)$ and then distribute them, round-robin, in halos
in order of decreasing luminosity.  This ensures that all the bright galaxies
are partitioned among the halos rather than having pairs end up in any one
halo.  A slightly different approach, which has very similar clustering
properties, was suggested by Yang et al.~(\cite{YMvdBC}).  Here we compute
$L_1$ such that a halo of mass $M$ has (on average) only 1 galaxy brighter
than $L_1$.  We then draw luminosities for the galaxies in this halo, allowing
only the brightest galaxy to have $L\ge L_1$.  This also suppresses the
higher moments of $\Phi(L|M)$ for bright galaxies.
We shall follow Yang et al.~(\cite{YMvdBC}) unless stated otherwise.


\begin{thebibliography}{99}
\bibitem[2003]{WMAP}
  Bennett C.L., et al., 2003, \apj, in press [astro-ph/0302207]
\bibitem[2000]{BCFBL}
  Benson A.J., et al., 2000, \mnras, 311, 793
\bibitem[2002]{BW02}
  Berlind A., Weinberg D.H., 2002, \apj, 575, 587 [astro-ph/0109001]
\bibitem[2003]{coil}
	Coil A., et al. {\em (the DEEP2 Team)},
	2003, \apj, submitted  [astro-ph/0305586]
\bibitem[1989]{ColKai}
  Cole S., Kaiser N., 1989, \mnras, 237, 1127
\bibitem[1994]{CAFNZ}
  Cole S., Aragon-Salamanca A., Frenk C.S., Navarro J.F., Zepf S.E., 1994,
    \mnras, 271, 781
\bibitem[2001]{2dF}
  Colless M.\ M., et al. {\em (the 2dFGRS Team)},
        2001, \mnras, 328, 1039 [astro-ph/0106498]
\bibitem[2002]{CooShe}
  Cooray A., Sheth R., 2002, Physics Reports, 372, 1   [astro-ph/0206508]
\bibitem[1985]{DEFW}
  Davis M., Efstathiou G., Frenk C.S., White S.D.M., 1985, \apj, 292, 371
\bibitem[2002]{DEEP2}
  Davis M., et al., {\em (the DEEP2 Team)}
        2002, SPIE, [astro-ph/0209419]
\bibitem[2002]{depropris}
	De Propris R., Driver S.\ P.,
	2002, JENAM2002 Galaxy Evolution Workshop  [astro-ph/0212520]
\bibitem[1980]{dres}
	Dressler A.
	1980, \apj, 236, 351
\bibitem[1988]{eep}
	Efstathiou G., Ellis R.S., Peterson B.A., 
        1988, \mnras, 232, 431
\bibitem[1988]{EFWD}
  Efstathiou G., Frenk C.S., White S.D.M., Davis M., 1988, \mnras, 235, 715
\bibitem[2001]{GKHW}
  Gardner, J.P., Katz, N., Hernquist, L., Weinberg, D.H., 2001, ApJ, 559, 131
\bibitem[1996]{gilk}
	 Gilks W.R., Richarson S., Spiegelhalter D.J., 1996,
  Markov Chain Monte Carlo in practice
  (London: Chapman and Hall).
\bibitem[2003]{GuzSel}
  Guzik J., Seljak U., 2003, \mnras in press [astro-ph/0201448]
\bibitem[2003]{hawkins}
	Hawkins E., et al., {\em (the 2dFGRS Team)}
	2003, \mnras, submitted  [astro-ph/0212375]
\bibitem[1998]{JinMoBor}
  Jing Y.P., Mo H.J., Borner G., 1998, \apj, 494, 1 [astro-ph/9708115]
\bibitem[1993]{KauWhiGui}
  Kauffmann G., White S.D.M., Guiderdoni B., 1993, \mnras, 264, 201
\bibitem[1999]{KatHerWei}
  Katz N., Hernquist L., Weinberg D.H., 1999, ApJ, 523, 463 [astro-ph/9806257]
\bibitem[2002]{lahav} 
        Lahav O., et al. {\em (the 2dFGRS Team)}, 2002, 
   mnras, 333, 961 [astro-ph/0112162]
\bibitem[1999]{lefevre}
        Le Fevre, O., et al., 1999, ASP, Vol. 176, 250 
\bibitem[2000]{MaFry}
        Ma C.-P., Fry J.\ N.,
        2000, \apj, 543, 503
\bibitem[1990]{maddoxa}
	Maddox S.\ J., Efstathiou G., Sutherland W.\ J., Loveday J.,
	1990, \mnras, 243, 692
\bibitem[1990]{maddoxb}
	Maddox S.\ J., Efstathiou G., Sutherland W.\ J.,
	1990, \mnras, 246, 433
\bibitem[2002]{DSMLF}
  	Madgwick D.\ S. et al.,  {\em (the 2dFGRS Team)} 2002, \mnras,
        333, 133 [astro-ph/0107197]
\bibitem[2003]{mad03}
  	Madgwick D.\ S., 2003, \mnras, 338, 197 [astro-ph/0209051]
\bibitem[2003a]{mad03a}
        Madgwick, D. S., Somerville, R., Lahav, O., Ellis, R. S.,
        2003a, \mnras, in press [astro-ph/0210471]
\bibitem[2003b]{mad03b}
  	Madgwick D.S. et al., {\em (the 2dFGRS Team)}
	2003b, \mnras, in press [astro-ph/0303668]
\bibitem[2003c]{mad03c}
  	Madgwick D.S. et al., {\em (the DEEP2 Team)}
	2003c, \apj, submitted [astro-ph/0305587]
\bibitem[1996]{MoWhi}
  Mo H-J., White S.D.M., 1996, \mnras, 282, 347
\bibitem[2001]{Norberg01}
  Norberg P., et al., 2001, \mnras, 328, 64 [astro-ph/0105500]
\bibitem[2002]{Norberg02}
  Norberg P., et al., 2002, \mnras, 332, 827 [astro-ph/0112043]
\bibitem[2000]{PeaSmi}
  	Peacock J.A., Smith R.E., 
	2000, \mnras, 318, 1144 [astro-ph/0005010]
\bibitem[1999]{Pearce}
  Pearce, F.R. et al., 1999, ApJ, 521, 99 [astro-ph/9905160]
\bibitem[2003]{MagPor}
  Maggliochetti M., Porciani C., 2003, \mnras submitted [astro-ph/0304003]
\bibitem[1974]{PS}
  Press W.H., Schechter P., 1974, ApJ, 187, 452
\bibitem[2001]{ScoShe}
  Scoccimarro R., Sheth R., 
  2001, \mnras, 329, 629 [astro-ph/0106120]
\bibitem[2001]{SSHJ}
  Scoccimarro R., Sheth R., Hui L., \& Jain B., 2001, ApJ, 546, 20
  [astro-ph/0006319]
\bibitem[2002]{Scr02}
  Scranton R., 2002, \mnras, 332, 697 [0108266]
\bibitem[2000]{Sel00}
  Seljak U., 2000, MNRAS, 318, 203  [astro-ph/0001493]
\bibitem[1999]{SheTor}
  Sheth R., Tormen G., 1999, \mnras, 308, 119 [astro-ph/9901122]
\bibitem[2003]{Smith}
  Smith R.E., et al., 2003, 
	\mnras, 341, 1311 [astro-ph/0207664]
\bibitem[1999]{SomPri}
  Somerville R., Primack J., 1999, \mnras, 310, 1087 [astro-ph/9802268]
\bibitem[2002]{Strauss}
  Strauss M.\ A., et al. {\em (the SDSS collaboration)},
        2002, AJ, 124, 1810 [astro-ph/0206225]
\bibitem[2002]{verde} 
        Verde L., et al. {\em (the 2dFGRS Team)}, 
        2002, \mnras, 335, 432 [0112161]
\bibitem[2001]{Whi01}
  White M., 2001, \mnras, 321, 1   [astro-ph/0005085]
\bibitem[2001]{WhiHerSpr}
  White M., Hernquist L., Springel V., 2001, \apj, 550, L129  [astro-ph/0012518]
\bibitem[1978]{WhiRee}
  White S.D.M., Rees M., 1978, \mnras, 183, 341
\bibitem[2002]{TreePM}
  White M., 2002, \apjs, 579, 16 [astro-ph/0207185]
\bibitem[2003]{YMvdB}
  Yang X., Mo H-J., van den Bosch F.C., 2003, 
  \mnras, 339, 1057 [astro-ph/0207019]
\bibitem[2003]{YMvdBC}
  Yang X., Mo H-J., van den Bosch F.C., Chu Y., 2003, \mnras submitted
    [astro-ph/0303524]
\bibitem[2001]{YTJS}
  Yoshikawa K., Taruya A., Jing Y.P., Suto Y., 2001, \apj, 558, 520
\bibitem[2003]{Zehavi03}
  Zehavi I., Weinberg D.H., Zheng Z., Berlind A.A., Frieman J.A., et al.,
  2003, \apj submitted [astro-ph/0301280]
\end{thebibliography}
\end{document}